\documentclass[twocolumn,showpacs,preprintnumbers,amsmath,amssymb]{revtex4}
\usepackage{graphicx}
\usepackage{dcolumn}
\usepackage{bm}
\begin{document}
%
\newcommand{\kvec}{\mbox{{\scriptsize {\bf k}}}}
\newcommand{\lvec}{\mbox{{\scriptsize {\bf l}}}}
\newcommand{\qvec}{\mbox{{\scriptsize {\bf q}}}}
%
\def\eq#1{(\ref{#1})}
\def\fig#1{\hspace{1mm}\ref{#1}}
\def\tab#1{\hspace{1mm}\ref{#1}}
\title{Anisotropic superconductivity in the two-dimensional electron-phonon system}
\author{R. Szcz{\c{e}}{\`s}niak}
\email{szczesni@wip.pcz.pl}
\affiliation{Institute of Physics, Cz{\c{e}}stochowa University of Technology, Al. Armii Krajowej 19, 42-200 Cz{\c{e}}stochowa, Poland}
\date{\today}
\begin{abstract}
The properties of the $d$-wave superconducting state in the two-dimensional system have been studied. It has been assumed, that the pairing mechanism is based on the electron-phonon and the electron-electron-phonon interactions. The obtained results have shown the energy gap amplitude ($\Delta_{tot}$) crossover, from the BCS to non-BCS behavior, as the value of the electron-electron-phonon potential increases. The model has been tested for the ${\rm La_{2-x}Sr_{x}CuO_{4}}$ and ${\rm Bi_{2}Sr_{2}CaCu_{2}O_{8+\delta}}$ high-$T_{C}$ superconductors. It has been shown, that the dependence of the  $2\Delta^{\left(0\right)}_{tot}/k_{B}T_{C}$ ratio on the hole density is in agreement with the experimental data.
\end{abstract}
\pacs{74.20.-z, 74.20.Fg, 74.20.Mn, 74.20.Rp, 74.25.Bt, 74.72.-h}
\maketitle


In year 1986, Bednorz and M{\"u}ller have discovered the high temperature superconductivity \cite{Bednorz}. Despite of the large efforts, the origin of the pairing correlations in high-$T_{C}$ materials is still unclear. In particular, the description of the cuprates' properties in the framework of the pure electronic models (the Hubbard or related approaches) is probably insufficient, since in these models the pairing correlations are too small \cite{Imada}. On the other hand, the conventional electron-phonon interaction is weak enough, that the depairing electronic correlations should completely suppress the phonon-mediated superconductivity \cite{Bohnen}.

In order to solve the discussed problem, we assume, that the pairing mechanism would inseparably link together the strong electron correlations and the crystal lattice vibrations (the electron-phonon (EPH) and electron-electron-phonon (EEPH) interaction). We notice, that the above idea has been originally postulated in \cite{Szczesniak1}, where the properties of the anomalous $s$-wave superconducting state have been studied. The obtained results have shown, that for the high value of the EEPH coupling, the $s$-wave energy gap weakly depends on the temperature ($T\leq T_{C}$), and above the critical temperature it vanishes at the Nernst temperature.   

In the presented paper we will study the thermodynamic properties of the $d$-wave superconducting state on the basis of the postulated pairing model. In particular, we will calculate the dependence of the energy gap amplitude $\Delta_{tot}$ on the temperature for the selected values of the EEPH potential. Next, we will analyze the dependence of the $2\Delta^{\left(0\right)}_{tot}/k_{B}T_{C}$ ratio on the hole density for 
${\rm La_{2-x}Sr_{x}CuO_{4}}$  (LSCO) and ${\rm Bi_{2}Sr_{2}CaCu_{2}O_{8+\delta}}$ (Bi2212) superconductors. The theoretical predictions will be compared with the experimental data.


The effective Hamiltonian takes the form \cite{Szczesniak1}:
\begin{equation}
\label{r1}
H\equiv H^{\left(0\right)}+H^{\left(1\right)}+H^{\left(2\right)}.
\end{equation}
The first term represents the non-interacting electrons:
\begin{equation}
\label{r2}
H^{\left(0\right)}\equiv\sum_{\kvec\sigma }\varepsilon _{\kvec}c_{\kvec\sigma}^{\dagger}c_{\kvec\sigma},
\end{equation}
where: $\varepsilon _{\kvec}=-t\gamma\left({\bf k}\right)$ denotes the electron band energy for the two-dimensional square lattice; $t$ is
the nearest-neighbor hopping integral and $\gamma\left({\bf k}\right)\equiv 2\left[\cos\left(k_{x}\right)+\cos\left(k_{y}\right)\right]$. The symbol $c^{\dagger}_{\kvec\sigma}$, ($c_{\kvec\sigma}$) is the creation (annihilation) operator for the electron with momentum ${\bf k}$ and spin $\sigma$. The EPH and EEPH interaction terms are given by:
\begin{equation}
\label{r3}
H^{\left(1\right)}\equiv\sum_{\kvec\qvec\sigma}V_{\kvec\qvec}
c^{\dagger}_{\kvec+\qvec-\sigma}c^{\dagger}_{-\kvec-\qvec\sigma}
c_{-\kvec\sigma}c_{\kvec-\sigma},
\end{equation}
and
\begin{equation}
\label{r4}
H^{\left(2\right)}\equiv\sum_{\kvec\kvec^{'}\qvec\lvec\sigma}
U_{\kvec\kvec^{'}\qvec\lvec}
c_{\kvec-\lvec\sigma }^{\dagger}c_{\kvec\sigma}
h_{\kvec^{'}\lvec\qvec\sigma}
c_{-\kvec+\lvec-\sigma }^{\dagger}c_{-\kvec-\sigma},
\end{equation}
where: $h_{\kvec^{'}\lvec\qvec\sigma}\equiv c_{\kvec^{'}+\lvec+\qvec-\sigma }^{\dagger}c_{\kvec^{'}-\sigma}
c_{-\kvec^{'}-\lvec-\qvec\sigma }^{\dagger}c_{-\kvec^{'}\sigma}$. The functions $V_{\kvec\qvec}$ and $U_{\kvec\kvec^{'}\qvec\lvec}$ indicate the pairing potentials:
\begin{equation}
\label{r5}
V_{\kvec\qvec}\equiv\frac{\omega_{0}|g^{\left(1\right)}|^{2}}
{\left(\varepsilon_{\kvec}-\varepsilon_{\kvec+\qvec}\right)^{2}-\omega^{2}_{0}},
\end{equation}
and
\begin{equation}
\label{r6}
U_{\kvec\kvec^{'}\qvec\lvec}\equiv
\frac{\omega_{0}|g^{\left(2\right)}|^{2}}
{\left(\varepsilon_{\kvec}-\varepsilon_{\kvec-\lvec}+\varepsilon_{\kvec^{'}}-
\varepsilon_{\kvec^{'}+\lvec+\qvec}\right)^{2}-\omega^{2}_{0}},
\end{equation}
where $g^{\left(1\right)}$ and $g^{\left(2\right)}$ represents the EPH and EEPH coupling respectively; $\omega_{0}$ denotes the characteristic phonon frequency.

On the basis of the operators \eq{r3} and \eq{r4} it is possible to deduce the Hamiltonians which describe the $d$-wave superconducting state. In the case of the Hamiltonian \eq{r3} this procedure is known and widely described in the literature (see e.g. the paper \cite{Newns}). With reference to the above, we will discuss only the derivation of the anisotropic Hamiltonian on the basis of the EEPH operator.
In the first step, we separate the momentums in the expression \eq{r4}:
\begin{equation}
\label{r7}
H^{\left(2\right)}\simeq
\sum_{\kvec_{1}\sim\kvec_{4}\sigma}U_{\kvec_{1}\sim\kvec_{4}}
c_{-\kvec_{1}\sigma }^{\dagger}c_{-\kvec_{2}\sigma}
h_{\kvec_{3}\kvec_{4}\sigma}
c_{\kvec_{1}-\sigma}^{\dagger}c_{\kvec_{2}-\sigma},
\end{equation}
where: $h_{\kvec_{3}\kvec_{4}\sigma}\equiv c_{\kvec_{3}-\sigma}^{\dagger}c_{-\kvec_{3}\sigma}^{\dagger}
c_{-\kvec_{4}\sigma}c_{\kvec_{4}-\sigma}$. 
%
\begin{figure}[t]%
\includegraphics*[width=\columnwidth]{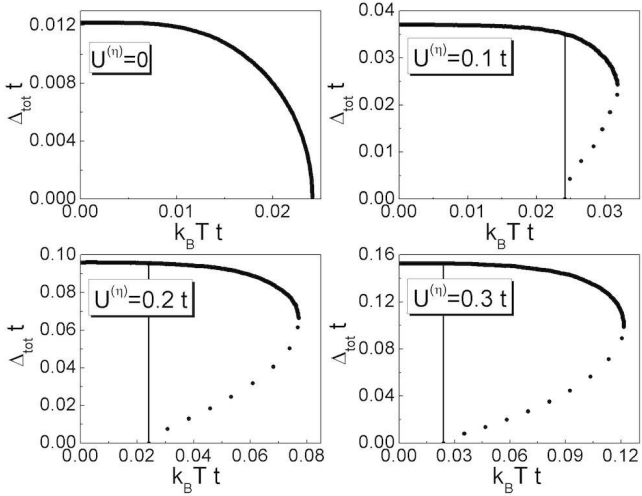}
\caption{The dependence of the energy gap amplitude on the temperature for the selected values of the EEPH potential. The solid line represents the physical stable solution; the dotted line corresponds to the unstable solution, where the thermodynamic potential is bigger than in the first case. The vertical line indicates the position of the critical temperature.}
\label{f1}
\end{figure}
%
With help of the relation: $c_{\kvec\sigma}=\frac{1}{\sqrt{N}}\sum_{j}e^{-i\kvec {\bf R}_{j}}c_{j\sigma}$, we transform the operator \eq{r7} to the Wannier representation, where we restrict ourselves to on-site and the nearest neighbor pairing. The Hamiltonian takes the form:
\begin{eqnarray}
\label{r8}
H^{\left(2\right)}&\simeq&
\sum_{\left<j_{1}j_{2}\right>\left<j_{3}j_{4}\right>\sigma}U^{\left(-\right)}_{j_{1}\sim j_{4}}
c_{j_{1}\sigma}^{\dagger}c_{j_{1}\sigma}
h^{\left(-\right)}_{j_{3}j_{4}\sigma}
c_{j_{2}-\sigma}^{\dagger}c_{j_{2}-\sigma}\\ \nonumber
&+&
\sum_{\left<j_{1}j_{2}\right>\left<j_{3}j_{4}\right>\sigma}U^{\left(+\right)}_{j_{1}\sim j_{4}}
c_{j_{1}\sigma}^{\dagger}c_{j_{2}\sigma}
h^{\left(+\right)}_{j_{3}j_{4}\sigma}
c_{j_{2}-\sigma}^{\dagger}c_{j_{1}-\sigma},
\end{eqnarray}
where 
$h^{\left(-\right)}_{j_{3}j_{4}\sigma}\equiv c_{j_{3}-\sigma}^{\dagger}c_{j_{3}-\sigma}c_{j_{4}\sigma}^{\dagger}c_{j_{4}\sigma}$ and  
$h^{\left(+\right)}_{j_{3}j_{4}\sigma}\equiv c_{j_{3}-\sigma}^{\dagger}c_{j_{4}-\sigma}c_{j_{4}\sigma}^{\dagger}c_{j_{3}\sigma}$.
The symbols $U^{\left(-\right)}_{j_{1}\sim j_{4}}$ and $U^{\left(+\right)}_{j_{1}\sim j_{4}}$ denote the local and kinetic potential respectively.
Next, we return to the Bloch representation: 
\begin{eqnarray}
\label{r9}
H^{\left(2\right)}&\simeq&
\sum_{\kvec_{1}\sim\kvec_{4}\sigma}
\left[U^{\left(-\right)}_{\kvec_{1}\sim\kvec_{4}}+U^{\left(+\right)}_{\kvec_{1}\sim\kvec_{4}}\right]\\ \nonumber
&\times&
c_{-\kvec_{1}\sigma }^{\dagger}c_{-\kvec_{2}\sigma}
h_{\kvec_{3}\kvec_{4}\sigma}
c_{\kvec_{1}-\sigma}^{\dagger}c_{\kvec_{2}-\sigma},
\end{eqnarray}
where:
\begin{eqnarray}
\label{r10}
U^{\left(\pm\right)}\left({\bf k}_{1}\sim {\bf k}_{4}\right)&\equiv& \frac{1}{N^{4}}
\sum_{\left<j_{1}j_{2}\right>}\sum_{\left<j_{3}j_{4}\right>}U^{\left(\pm\right)}_{j_{1}\sim j_{4}}\\ \nonumber
&\times&
e^{\mp i\left(\kvec_{2}\pm\kvec_{1}\right)\left({\bf R}_{j_{2}}-{\bf R}_{j_{1}}\right)\pm i\left(\kvec_{4}\pm\kvec_{3}\right)\left({\bf R}_{j_{4}}-{\bf R}_{j_{3}}\right)}.
\end{eqnarray}
Now, we assume: $U^{\left(-\right)}_{j_{1}\sim j_{4}}\simeq U^{\left(+\right)}_{j_{1}\sim j_{4}}=-U/24N_{0}^{3}$. 
We notice that, the number $24$ has  been introduced, since the potential energy term represents the interaction between every four of particles counted once; $N_{0}$ is the normalization factor: $N_{0}\equiv 1/\sum^{\omega_{0}}_{\kvec}$, where the symbol $\sum^{\omega_{0}}_{\kvec}$ denotes the sum over the states for which $|\varepsilon_{\kvec}|\leq\omega_{0}$ (the BCS-type cut-off). In the next step, we limit the symmetry of the energy gap to the dominating $d$-wave symmetry. The total Hamiltonian after applying the approximation presented in the paper \cite{Szczesniak1} takes the form: 
\begin{eqnarray}
\label{r11}
H^{\left(\eta\right)}&\equiv&
\sum_{\kvec\sigma}\varepsilon_{\kvec}c^{\dagger}_{\kvec\sigma}c_{\kvec\sigma}
-\left(V^{\left(\eta\right)}+\frac{U^{\left(\eta\right)}}{6}\left|\Delta^{\left(\eta\right)}\right|^{2}\right)\\ \nonumber
&\times&
\sum^{\omega_{0}}_{\kvec}\left[
\Delta^{\left(\eta\right)}_{\kvec}c_{\kvec\uparrow }^{\dagger}c_{-\kvec\downarrow}^{\dagger}+
\Delta^{\star\left(\eta\right)}_{\kvec}c_{-\kvec\downarrow }c_{\kvec\uparrow}\right].
\end{eqnarray}
The symbol $V^{\left(\eta\right)}$ and $U^{\left(\eta\right)}$ denotes the $d$-wave effective potential for the EPH and EEPH channel respectively. In particular: $V^{\left(\eta\right)}\equiv V/2$ and $U^{\left(\eta\right)}\equiv U/8$. The anisotropic order parameter is given by:   $\Delta^{\left(\eta\right)}_{\kvec}\equiv\Delta^{\left(\eta\right)}\eta\left(\bf{k}\right)$, where the amplitude is expressed as:
 $\Delta^{\left(\eta\right)}\equiv\frac{1}{N_{0}}\sum^{\omega_{0}}_{\kvec}\eta\left({\bf k}\right)\left<c_{-\kvec\downarrow}c_{\kvec\uparrow}\right>$ and $\eta\left({\bf k}\right)\equiv 2\left[\cos\left(k_{x}\right)-\cos\left(k_{y}\right)\right]$.

On the basis of the Hamiltonian \eq{r11}, we calculate the thermodynamic Green function by using the equation of motion method \cite{Gasser}. The result has the form:
\begin{equation}
\label{r12}
\left<\left<c_{\kvec\uparrow}|c_{-\kvec\downarrow}\right>\right>=-
\frac{\left(V^{\left(\eta\right)}+\frac{U^{\left(\eta\right)}}{6}\left|\Delta^{\left(\eta\right)}\right|^{2}\right)\Delta^{\left(\eta\right)}_{\kvec}}
{\omega^{2}-\left(E^{\left(\eta\right)}_{\kvec}\right)^{2}},
\end{equation} 
where: 
\begin{equation}
\label{r13}
E^{\left(\eta\right)}_{\kvec}\equiv\sqrt{\epsilon^{2}_{\kvec}+\left(V^{\left(\eta\right)}
+\frac{U^{\left(\eta\right)}}{6}\left|\Delta^{\left(\eta\right)}\right|^{2}\right)^{2}
\left(\left|\Delta^{\left(\eta\right)}\right|\eta\left({\bf k}\right)\right)^{2}}.
\end{equation}
We turn the reader's attention toward the fact, that the obtained Green function possesses the analytical structure, which is more complex than the structure of the BCS Green function \cite{Gasser}, \cite{Bardeen}. In particular, the energy gap amplitude is the complicated function of the order parameter amplitude. However, the energy gap, in spite of its complicated form, is characterized by the pure $d$-wave symmetry.

On the basis of Eq.\eq{r12} we derive the fundamental thermodynamic equation:
\begin{equation}
\label{r14}
1=\left(V^{\left(\eta\right)}+\frac{U^{\left(\eta\right)}}{6}\left|\Delta^{\left(\eta\right)}\right|^{2}\right)
\frac{1}{N_{0}}\sum^{\omega_{0}}_{\kvec}\frac{\eta^{2}\left({\bf k}\right)}{2E^{\left(\eta\right)}_{\kvec}}\tanh\frac{\beta E^{\left(\eta\right)}_{\kvec}}{2}.
\end{equation} 
The sum $\sum^{\omega_{0}}_{\kvec}$ is approximated in the following manner:  
$\sum^{\omega_{0}}_{\kvec}\simeq \int^{\pi}_{-\pi}\int^{\pi}_{-\pi}dk_{x}dk_{y}\theta\left(\omega_{0}-|\varepsilon_{k_{x},k_{y}}|\right)$, where the symbol $\theta$ represents the unit step function. In the model calculations, we have taken $t$ as the energy unit.


%
\begin{table*}
\caption{\label{t1} The parameters $V^{\left(\eta\right)}$ and $U^{\left(\eta\right)}$ calculated by using $T_{C}$ and the mean values of $T^{*}$.}
\begin{ruledtabular}
\begin{tabular}{ccccccccccc}
Material & Type &$t$ (meV)&Ref.&$\omega_{0}$ (meV)&Ref.&$T_{C}$ (K)&$T^{*}$ (K)&Ref.&$V^{\left(\eta\right)}$ (meV)&$U^{\left(\eta\right)}$ (meV)\\
         &      &         &    &                  &    &           &           &    &                             &                             \\
\hline
LSCO    &p=0.10\footnote{The hole density $p$ has been estimated as the doping $p=x$.}      
                     &$240$&\cite{Xu}  &$96$&\cite{Kim1}                  &$25$         &$193.7\pm 30.2$&\cite{Hashimoto}&{\bf 3.76}&{\bf 66.86}\\
        &p=0.15      &     &           &    &                             &$38$         &$135.5\pm 30.2$&\cite{Hashimoto}&{\bf 4.41}&{\bf 53.89}\\
        &p=0.22      &     &           &    &                             &$28$         &$44.3 \pm 30.2$&\cite{Hashimoto}&{\bf 3.91}&{\bf 45.11}\\
\hline
Bi2212  &$p=0.125$\footnote{The hole density $p$ has been obtained on the basis of the empirical formula 
                            $T_{C}\left(p\right)/T_{C,{\rm max}}=1-82.6\left(p-0.16\right)^{2}$, 
                            where $T_{C,{\rm max}}=92.2$ K \cite{Presland}.} &$350$&\cite{Tohayama},\cite{Tohayama1},&$80$&\cite{Damascelli},\cite{Cuk},&$83$ &$\sim 290$ &\cite{Renner}&{\bf 5.47}&{\bf 55.40}\\
        &$p=0.143$ &     &\cite{Kim}.                      &    &\cite{Gweon},\cite{Kulic},  &$90$ &$\sim 300$ &\cite{Matsuda1}&{\bf 5.68}&{\bf 56.37}\\
        &$p=0.193$ &     &                                 &    &\cite{Gonnelli}.            &$84$ &$\sim 270$ &\cite{Matsuda1}&{\bf 5.51}&{\bf 51.72}\\
        &$p=0.198$ &     &                                 &    &                            &$81$ &$\sim 220$ &\cite{Matsuda1}&{\bf 5.40}&{\bf 43.88}\\
\end{tabular}
\end{ruledtabular}
\end{table*}

In Fig.\fig{f1} we present the temperature dependence of the energy gap amplitude 
($\Delta_{tot}\equiv \left(V^{\left(\eta\right)}+\frac{U^{\left(\eta\right)}}{6}
\left|\Delta^{\left(\eta\right)}\right|^{2}\right)\left|\Delta^{\left(\eta\right)}\right|$) for $V^{\left(\eta\right)}=0.02t$, $\omega_{0}=0.3t$ and the selected values of $U^{\left(\eta\right)}$. It is easy to see, that for the high values of the EEPH potential, the shape of the function $\Delta_{tot}\left(T\right)$ is sharply different from the BCS prediction. In particular, for $T\in\left<0,T_{C}\right>$ the energy gap is very weakly dependent on the temperature; up to the critical temperature $\Delta_{tot}$ extends into the anomalous normal state to the temperature $T^{*}$. In the case of the $d$-wave superconducting state, the temperature $T^{*}$ is interpreted as the pseudogap temperature (in contrast to the $s$-wave superconductivity, where the highest value of the temperature, for which the non-zero solution of the gap equation exists, is connected with the Nernst temperature $T^{**}$ \cite{Szczesniak1}). We notice, that usually: $T^{*}>T^{**}$.

Below we compare the theoretical predictions with the experimental data for LSCO and Bi2212 superconductors. For this purpose, we have calculated the values of the pairing potentials on the basis of $T_{C}$ and $T^{*}$ experimental values. The obtained results have been collected in Tab.\tab{t1}. Next, by using the $V^{\left(\eta\right)}$ and $U^{\left(\eta\right)}$ values, the hole density dependence of the  $R_{1}\equiv 2\Delta^{\left(0\right)}_{tot}/k_{B}T_{C}$ ratio has been obtained. We notice, that the energy gap amplitude at the temperature of zero Kelvin is defined as: $\Delta^{\left(0\right)}_{tot}\equiv\left(V^{\left(\eta\right)}+\frac{U^{\left(\eta\right)}}{6}
\left|\Delta^{\left(\eta\right)}_{0}\right|^{2}\right)\left|\Delta^{\left(\eta\right)}_{0}\right|$, where $\Delta^{\left(\eta\right)}_{0}$ denotes 
the order parameter amplitude at the temperature of zero Kelvin. 

%
\begin{figure}[t]%
\includegraphics*[width=\columnwidth]{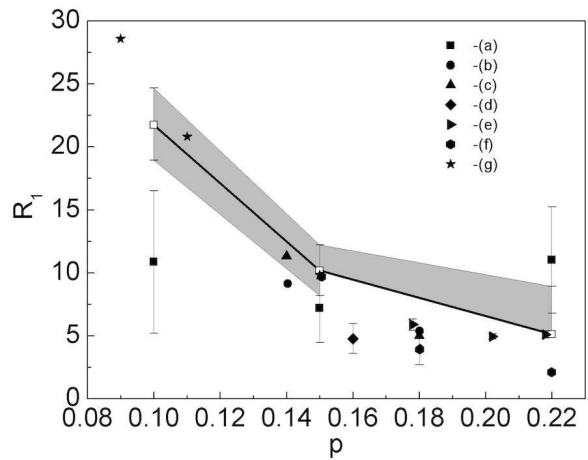}
\caption{The ratio $R_{1}$ as a function of $p$ for LCSO superconductor. The solid lines with the open squares represent the theoretical calculations based on the data presented in the paper \cite{Hashimoto}. The overshadowed areas mean the accuracy of the achieved results. The filled symbols correspond to the experimental results obtained by:
(a) - Hashimoto, {\it et al.} \cite{Hashimoto}, 
(b) - Nakano,    {\it et al.} \cite{Nakano},
(c) - Oda,       {\it et al.} \cite{Oda1}, 
(d) - Kato,    {\it et al.} \cite{Kato},
(e) - Wang,      {\it et al.} \cite{Wang}, 
(f) - Yoshida,   {\it et al.} \cite{Yoshida},
(g) - Wen,     {\it et al.} \cite{Wen}.}
\label{f2}
\end{figure}
%
\begin{figure}[t]%
\includegraphics*[width=\columnwidth]{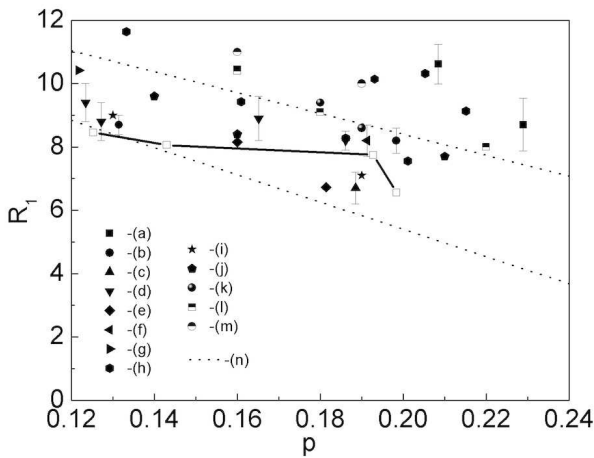}
\caption{The dependence of the ratio $R_{1}$ on $p$ for Bi2212. The solid line with the open squares represents the theoretical calculation. The filled and half-filled symbols correspond to the experimental results obtained by:
(a) - Renner, {\it et al.} \cite{Renner}, 
(b) - Hoffmann, {\it et al.} \cite{Hoffmann},
(c) - Ponomarev, {\it et al.} \cite{Ponomarev}, 
(d) - Oki, {\it et al.} \cite{Oki},
(e) - Krasnov, {\it et al.} \cite{Krasnov}, 
(f) - Gupta, {\it et al.} \cite{Gupta},
(g) - Kanigel, {\it et al.} \cite{Kanigel}, 
(h) - Campuzano, {\it et al.} \cite{Campuzano}, Tanaka, {\it et al.} \cite{Tanaka},
(i) - Nakano,  {\it et al.} \cite{Nakano},
(j) - Oda,     {\it et al.} \cite{Oda},
(k) - McElroy, {\it et al.} \cite{McElroy},
(l) - Matsuda, {\it et al.} \cite{Matsuda},
(m) - Hoffman, {\it et al.} \cite{Hoffman}.
The lines (n) were obtained by using the empirical relation: $R_{1}\left(p\right)=\left(15\pm 1\right)-\left(38\pm 5\right)p$ \cite{Hewitt}.}
\label{f3}
\end{figure}
%

In Fig.\fig{f2} we present the dependence of the $R_{1}$ ratio on the hole density for LSCO superconductor. 
It can be seen, that with the increase of $p$, the parameter $R_{1}$ successively decreases. In particular, for the underdoped region ($p<0.155$) the  
values of $R_{1}$ are significantly higher than the $d$-wave BCS value $4.28$ \cite{Won}. Slightly above $p=0.155$ (the overdoped region) the $R_{1}$ ratio approaches closely the weak-coupling $d$-wave BCS result. We notice, that for $p>0.15$, the lower accuracy of the theoretical results can not be determined, since $T_{C}$ and $T^{*}$ are experimentally indistinguishable (see Tab. \tab{t1}). Now, we have compared the theoretical predictions which the experimental values of $R_{1}$, received by the few different researchers. The qualitative agreement of the theoretical predictions with the experimental data proves, that the measured dependence of the ratio $R_{1}$ on $p$ can be well reproduced with an use of the presented model. 

In Fig.\fig{f3} we show the shape of the function $R_{1}\left(p\right)$ for Bi2212 superconductor. The presented results prove, that the theoretical line determines the high value of $R_{1}$ in the whole range of the considered hole density; $p\in\left<0.125,198\right>$. Important is also the fact, that the model correctly reconstructs the experimental data.
 

To summarize, we have presented the model that describes the properties of the $d$-wave superconducting state in the two-dimensional system. We have assumed, that the pairing correlations are induced by the electron-phonon and electron-electron-phonon interactions. In the first step, we have derived the fundamental thermodynamic equation. Next, on the basis of the exact numerical solution, we have shown, that for the high value of the EEPH potential, the temperature dependence of the energy gap amplitude differs sharply from the BCS prediction. In particular, the energy gap amplitude is slightly dependent on the temperature for $T\in\left<0,T_{C}\right>$; above the critical temperature, the energy gap amplitude persists to the pseudogap temperature. 
In the paper the theoretical predictions have been compared with the experimental data for LCSO and Bi2212 superconductors. It has been shown, that the calculated hole density dependence of the $R_{1}$ ratio correctly reproduces the experimental results. 

\begin{acknowledgments}
The author wish to thank Prof. K. Dzili{\'n}ski, the Head of the Institute of Physics at Cz{\c{e}}stochowa University of Technology, for providing excellent working conditions and the financial support. Additionally, I would like to thank my colleagues: D. Szcz{\c{e}}{\'s}niak, M.W. Jarosik and A.P. Durajski for their kindness and technical support given during the preparation of this paper. Some computational resources have been provided by the RSC Computing Center.
\end{acknowledgments}
%
%
 
%

\begin{thebibliography}{[2]}
%
\bibitem{Bednorz}
J.G. Bednorz, K.A. M{\"u}ller, Z. Phys. B {\bf 64}, 189 (1986).
\bibitem{Imada}
(a) M. Imada, Y. Hatsugai, J. Phys. Soc. Jpn. {\bf 58}, 3752 (1989);
(b) M. Imada, J. Phys. Soc. Jpn. {\bf 60}, 2740 (1991);
(c) D.J. Scalapino, S.R. White, S.C. Zhang, Phys. Rev. Lett. {\bf 68}, 2830 (1992);
(d) J.E. Hirsch, in: Proceedings of the International Conference on Strongly Correlated Electron Systems, San Diego,
August 1993.
\bibitem{Bohnen}
K.-P. Bohnen, R. Heid, M. Krauss, Europhys. Lett. 64, 104 (2003).
\bibitem{Szczesniak1}
R. Szcz{\c{e}}{\'s}niak, arXiv:1105.5525.
\bibitem{Newns}
D.M. Newns, C.C. Tsuei, P.C. Pattnaik, Phys. Rev. {\bf 52}, 13611 (1995).
\bibitem{Gasser}
W. Gasser, E. Heiner, K. Elk, Greensche Funktionen in Festk{\"o}rper- und Vielteilchenphysik, VILEY-VCH Verlag GmbH,
Weinheim (1999).
\bibitem{Bardeen}
(a) J. Bardeen, L.N. Cooper, J.R. Schrieffer, Phys. Rev. {\bf 106}, 162 (1957);
(b) J. Bardeen, L.N. Cooper, J.R. Schrieffer, Phys. Rev. {\bf 108}, 1175 (1957).
\bibitem{Xu}
J.H. Xu, T.J. Watson-Yang, J. Yu, A.J. Freeman, Phys. Lett. {\bf 120A}, 489 (1987).
\bibitem{Kim1}
J.H. Kim, Z. Tesanovic, Phys. Rev. Lett. {\bf 71}, 4218 (1993).
\bibitem{Hashimoto}
M. Hashimoto, T. Yoshida, K. Tanaka, A. Fujimori, M. Okusawa, 
S. Wakimoto, K. Yamada, T. Kakeshita, H. Eisaki, S. Uchida, Phys. Rev. B (R) {\bf 75}, 140503 (2007).
\bibitem{Presland}
M.R. Presland, J.L. Tallon, R.G. Buckley, R.S. Liu, N.E. Flower, Physica C {\bf 176}, 95 (1991).
\bibitem{Tohayama}
T. Tohayama, S. Maekawa, Supercond. Sci. Technol. {\bf 13}, R17 (2000).
\bibitem{Tohayama1}
T. Tohayama, S. Maekawa, Phys. Rev. B {\bf 67}, 092509 (2003).
\bibitem{Kim}
C. Kim, P.J. White, Z.-X. Shen, T. Tohyama, Y. Shibata, S. Maekawa, B.O. Wells, Y.J. Kim, R.J. Birgeneau, M.A. Kastner, 
Phys. Rev. Lett. {\bf 80}, 4245 (1998).
\bibitem{Damascelli}
A. Damascelli, Z. Hussain, Z.-X. Shen, Rev. Mod. Phys. {\bf 75}, 473 (2003).
\bibitem{Cuk}
T. Cuk, D.H. Lu, X.J. Zhou, Z.-X. Shen, T.P. Deveraux, N. Nagaosa, Phys. Stat. Sol. (b) {\bf 242}, 11 (2005).
\bibitem{Gweon}
G.-H. Gweon, T. Sasagawa, S.Y. Zhou, J. Graf, H. Takagi, D.-H. Lee , A. Lanzara, Nature {\bf 430}, 187 (2004).
\bibitem{Kulic}
M.L. Kulic, O.V. Dolgov, Phys. Rev. B {\bf 76}, 132511 (2007).
\bibitem{Gonnelli}
R.S. Gonnelli, G.A. Ummarino, V.A. Stepanov, Physica C {\bf 275}, 162 (1997).
\bibitem{Renner}
(a) Ch. Renner, B. Revaz, J.-Y. Genoud, K. Kadowaki, O. Fischer, Phys. Rev. Lett. {\bf 80}, 149 (1998);
(b) Ch. Renner, B. Revaz, K. Kadowaki, I. Maggio-Aprile, O. Fischer, Phys. Rev. Lett. {\bf 80}, 3606 (1998).
\bibitem{Matsuda1}
A. Matsuda, S. Sugita, T. Watanabe, Phys. Rev. B {\bf 60}, 1377 (1999).
\bibitem{Nakano}
T. Nakano, N. Momono, M. Oda, M. Ido, J. Phys. Soc. Jpn. {\bf 67}, 2622 (1998).
\bibitem{Oda1}
M. Oda, T. Matsuzaki, N. Momono, M. Ido, Physica C {\bf 341}, 847 (2000).
\bibitem{Kato}
T. Kato, H. Morimoto, A. Katagiri,S. Okitsu, H. Sakata, Physica C {\bf 392}, 221 (2003).
\bibitem{Wang}
Y. Wang, J. Yan, L. Shan, H.-H. Wen, Y. Tanabe, T. Adachi, Y. Koike, Phys. Rev. B {\bf 76}, 064512 (2007).
\bibitem{Yoshida}
T. Yoshida, X.J. Zhou, D.H. Lu, S. Komiya, Y. Ando, H. Eisaki, T. Kakeshita, 
S. Uchida, Z. Hussain, Z.-X. Shen, A. Fujimori, J. Phys.: Condens. Matter {\bf 19}, 125209 (2007).
\bibitem{Wen}
H.H. Wen, L. Shan, X.G. Wen, Y. Wang, H. Gao, Z.Y. Liu, 
F. Zhou, J.W. Xiong, W.X. Ti, Phys. Rev. B {\bf 72}, 134507 (2005).
\bibitem{Won}
H. Won, K. Maki, Phys. Rev. B {\bf 49}, 1397 (1994).
\bibitem{Hoffmann}
A. Hoffmann, P. Lemmens, L. Winkeler, G. Guntherodt, J. Low Temp. Phys. {\bf 99}, 201 (1995).
\bibitem{Ponomarev}
Y.G. Ponomarev, N.Z. Timergaleev, A.O. Zabezhaylov, K.K. Uk, M.A. Lorenz, G. Muller, H. Piel, H. Schmidt, C.
Janowitz, A. Krapf, R Manzke, Conference Series-Institute of Physics, {\bf 2}, 167 (2000).
\bibitem{Oki}
T. Oki, N. Tsuda, D. Shimada, Physica C {\bf 353}, 213 (2001).
\bibitem{Krasnov}
V.M. Krasnov, A. Yurgens, D. Winkler, P. Delsing, T. Claeson, Phys. Rev. Lett. {\bf 84}, 5860 (2000).
\bibitem{Gupta}
A.K. Gupta, K.-W. Ng, Phys. Rev. B {\bf 58},  R8901 (1998).
\bibitem{Kanigel}
A. Kanigel, U. Chatterjee, M. Randeria, M.R. Norman, S. Souma, 
M. Shi, Z.Z. Li, H. Raffy, J.C. Campuzano, Phys. Rev. Lett. {\bf 99}, 157001 (2007).
\bibitem{Campuzano}
J.C. Campuzano, H. Ding, M.R. Norman, H.M. Fretwell, M. Randeria, A. Kaminski, J. Mesot, T. Takeuchi, T. Sato,
T. Yokoya, T. Takahashi, T. Mochiku, K. Kadowaki, P. Guptasarma, D.G. Hinks, Z. Konstantinovic, Z.Z. Li, H. Raffy,
Phys. Rev. Lett. {\bf 83}, 3709 (1999).
\bibitem{Tanaka}
K. Tanaka, W.S. Lee, D.H. Lu, A. Fujimori, T. Fujii, Risdiana, I. Terasaki, D.J. Scalapino, T.P. Devereaux, Z. Hussain,
Z.-X. Shen, Science {\bf 314}, 1910 (2006).
\bibitem{Oda}
M. Oda, K. Hoya, R. Kubota, C. Manabe, N. Momono, T. Nakano, M. Ido, Physica C {\bf 281}, 135 (1997).
\bibitem{McElroy}
K. McElroy, D.-H. Lee, J.E. Hoffmann, K.M. Lang, J. Lee, E.W. Hudson, H. Eisaki, S. Uchida, J.C. Davis,
Phys. Rev. Lett. {\bf 94}, 197005 (2005).
\bibitem{Matsuda}
A. Matsuda, T. Fujii, T. Watanabe, Physica C {\bf 388-389}, 207 (2003).
\bibitem{Hoffman}
J.E. Hoffman, E.W. Hudson, K.M. Lang, V. Madhavan, H. Eisaki, S. Uchida, J.C. Davis, Science {\bf 295}, 466 (2002).
\bibitem{Hewitt}
K.C. Hewitt, J.C. Irwin, Phys. Rev. B {\bf 66}, 054516 (2002).
%
\end{thebibliography}
\end{document}